# Identification of a Distinct Bosonic Mode in an Electron-Doped Superconductor

F. C. Niestemski[1], S. Kunwar[1], S. Zhou[1], Shiliang Li[2], H. Ding[1], Ziqiang Wang[1], Pengcheng Dai[2, 3], V. Madhavan[1]

[1] *Department of Physics, Boston College, Chestnut Hill, Massachusetts 02467, USA*

[2] *Department of Physics and Astronomy, The University of Tennessee, Knoxville, Tennessee 37996-1200, USA*

[3] *Neutron Scattering Sciences Division, Oak Ridge National Laboratory, Oak Ridge, Tennessee 37831-6393, USA*

**Despite recent advances in understanding high-temperature (high-$T_c$) superconductors, there is no consensus on the origin of the superconducting 'glue' i.e., the mediator that binds electrons into superconducting pairs. The main contenders are lattice vibrations (phonons)[1,2] and spin excitations[3,4] with the additional possibility of RVB pairing without mediators[5]. In conventional superconductors, phonon mediated pairing was unequivocally established by data from tunneling experiments[6]. Proponents of phonons as the high-$T_c$ glue were therefore reinvigorated by the recent scanning tunneling microscopy (STM) experiments on hole-doped $Bi_2Sr_2CaCu_2O_{8-\delta}$ (BSCCO)[7] that reveal an oxygen lattice vibrational mode intimately connected to the superconducting gap energy scale. Here we report the first high resolution STM measurements of the electron-doped high-$T_c$ superconductor $Pr_{0.88}LaCe_{0.12}CuO_4$ (PLCCO) ($T_c$ = 24 K) revealing a bosonic excitation (mode) at energies 10.5 ± 2.5 meV. This energy is consistent with both spin-excitations in PLCCO measured by the neutron resonance mode[8] and a low energy acoustic phonon mode[9] but differs substantially from the oxygen vibrational mode identified in BSCCO. Our analysis of the variation of the local mode energy and intensity with the local gap energy scale indicates an electronic origin of the mode consistent with spin-excitations rather than phonons. These new findings place significant constraints on pairing theories that seek to encompass both these classes of high-$T_c$ materials.**

Electron- and hole-doped high-$T_c$ superconductors share identical $CuO_2$ planes where superconductivity originates. Compared to their hole-doped counterparts, the electron-doped cuprates represent a largely unexplored territory for STM studies where the lack of high quality samples has posed a tremendous barrier to obtaining high quality data comparable to that on BSCCO. We have obtained reproducible STM data on nearly optimally doped PLCCO ($T_c$ = 24 K) used in recent neutron scattering[8] and ARPES experiments[10]. Figure 1 shows selected STM spectra illustrating the most prominent features in the density of states (DOS): the superconducting gap with coherence peaks and the step/peak features outside the gap. In addition to these, an obvious feature of the tunneling spectra is the presence of an almost linear, V-shaped background (figure 1b, also see supplementary figure 2) which persists above $T_c$ (figure 2d). A similar background was observed in prior STM data on the electron-doped superconductor $Nd_{2-x}Ce_xCuO_4$ (NCCO)[11]. There are several different conjectures for a linear background in the DOS, ranging from a marginal Fermi liquid self-energy effects[12] to inelastic tunneling from a continuum of states[13]. Once this background is divided out however, the muted spectral features (including the formerly suppressed coherence peak heights) come to the forefront as shown in figure 1c and d. To make sure that the observed gap is associated with the superconducting gap, we have performed spectroscopy at various temperatures up to 32K (> $T_c$) (figure 2d) and find that the gap does indeed disappear above $T_c$. We thus identify the peak to peak distance in the local density of states (LDOS) at 5.5 K, with twice the local energy gap for superconducting quasiparticles (2Δ).

While there have been no prior STM studies on PLCCO, ARPES studies[14] point towards a non-monotonic *d*-wave gap with a maximum around 5.5 meV. Point contact tunneling[15] observes a zero temperature gap (Δ(0)) of 3.5 meV with a ratio *2Δ(0)/$k_BT_c$* =

3.5±0.3 consistent with weak-coupling BCS. Previous STM data obtained on NCCO[11] showed 3.5 meV - 5 meV gaps with no obvious coherence peaks. Given the highly inhomogeneous nature of doped layered oxides, spatially resolved STM is a useful key to providing the local energy scales and spatial distribution of the superconducting gap. Statistics of the gap magnitude and its spatial variation were obtained through thousands of spectra (*dI/dV* mapping) in various regions of the sample (figure 2a and 2c). While most maps (9 out of 13) reveal average gaps in the range of 6.5-7.0 meV, the average gap (over all measured maps) is 7.2 ± 1.2 meV (figure 2c). Approximating $\Delta(0)$ as 7.2 meV allows us to obtain a rough estimate for the ratio *2Δ/k$_B$T$_c$* ~ 7.5, putting the electron-doped superconductors in the strong coupling regime, thereby suggesting a greater overlap between the fundamental physics of the electron and hole-doped materials than previously shown.

We now turn to important features in the LDOS at energies greater than $\Delta$. A step-like feature in the DOS (which results in a peak in the second derivative of the tunnel current *d$^2$I/dV$^2$*) is normally interpreted as the signature of a bosonic excitation in the system. STM data on bosonic excitations and the strengths of their coupling to the electrons could potentially provide critical information on viable candidates for the pairing mode. Shown in figure 3a is a typical *dI/dV* spectrum obtained on these samples. The derivative of the spectrum (figure 3b) reveals peaks at distinct energies marked E$_1$ and E$_2$. In the superconducting state, a bosonic excitation appears in STM spectra at an energy offset by the gap, i.e. E = $\Omega$ + $\Delta$ where E is the energy of the feature in the spectrum and $\Omega$ is the mode energy. While the lineshape of the feature could be influenced by the details of the process, both inelastic tunneling effects[16] and electron self-energy effects[6] from a strongly coupled bosonic mode are expected at energies offset by the gap. This allows us to conveniently extract the mode energies ($\Omega_{1,2}$ = E$_{1,2}$ – 7.0) resulting in 10.7 meV and 21.7 meV respectively. Since spectral features at multiples of $\Omega_1$ could arise from multi-boson excitations, we find that $\Omega_{1,2}$ are amenable to interpretation as multiples of the same mode $\Omega_1$ at $10.85 \pm 0.15$ meV.

To determine the statistical significance of the observation of this bosonic mode, high resolution *dI/dV* maps were obtained over many regions of the samples and analyzed to extract $\Delta$ and E locally. Data from one such map is shown in figure 3c. The observation of multiples of $\Omega_1$ allows us to extract $\Omega_1$ in two different ways for each spectrum: $\Omega_1 = E_1-\Delta$ and $\Omega_1^* = E_2-E_1$. These are independent observables whose histograms are plotted in figure 3d. As can be seen, the two histograms overlap strongly. We thus conclude that the identification of $\Omega_2$ as $2\Omega_1$ bears significant statistical weight, which further supports the identification of the steps/peaks outside the superconducting gap as originating from bosonic excitations in PLCCO. Using the data from eight *dI/dV* maps (figure 3e) we obtain an average mode energy of $\Omega_{1av}= 10.5 \pm 2.5$ meV. It is worthwhile to note that both the intensity of $\Omega_1$ and the observation of the second harmonic ($2\Omega_1$) of the mode indicate a relatively strong electron-mode coupling. From these spatially resolved spectra, we can also calculate the correlation between the local mode $\Omega(r)$ and the gap $\Delta(r)$. We find that $\Omega(r)$ is anti-correlated with the local gap magnitude $\Delta(r)$ as visible in figure 4a. The obtained correlation function between the two is quite short-ranged, with a normalized on-site value close to -0.4, comparable to that found in BSCCO[7]. This anti-correlation is the first indication that this signal arises from an intrinsic excitation rather than an extrinsic inelastic excitation outside the superconducting planes.

Having established the mode energy, statistics and its correlation to the local gap, we now discuss the nature of this excitation. Indeed, the measured mode energy of 10.5±2.5meV suggests an immediate connection to the 11 meV magnetic resonance mode discovered recently in PLCCO[8] and NCCO[17] at Q= (½, ½, 0) by inelastic neutron scattering. The neutron resonance mode, or more precisely its precursor above $T_c$, has been suggested as a possible pairing glue for the high-$T_c$ cuprates. Theoretically, bosonic modes originating from spin-excitations can be observed by STM provided there is sufficient coupling between the charge and spin degrees of freedom[18]. Magnetoresistance measurements on underdoped non-superconducting $Pr_{1.3-x}La_{0.7}Ce_xCuO_{4-\delta}$ have provided evidence for strong spin-charge coupling in these materials[19]. It is thus possible that the magnetic resonance mode observed by neutron scattering is related to the observed bosonic mode in the STM signal in PLCCO.

While magnetic excitations fit the energy scale of our data, another possibility is that the mode originates from in-plane ($CuO_2$ plane) phonons, like the $B_{1g}$ mode attributed to the STM feature in BSCCO. Compared to BSCCO, however, the energy scale of our mode (10.5 ± 2.5meV) is much lower. In the hole-doped superconductors a few phonon branches do exist at these low energies[20, 21] and the important question is whether there are candidate phonons at these low energies in PLCCO. As it turns out many of the in-plane phonons in closely related materials, including the $B_{1g}$ mode, have energies higher than 20 meV[22, 23, 24] and can therefore be ruled out as possible candidates. Acoustic phonons are viable candidates for this mode provided the phonon dispersion results in a sharp DOS feature at this energy scale. Such phonons with a DOS peak at or close to 11 meV have indeed been found in NCCO[9, 25]. Expanding the search to in-plane phonons at nearby energies reveals a $E_u$ oxygen mode[23] and an oxygen rotation mode[22] at energies greater than 15 meV (15 meV is the lowest energy in the dispersion). We thus conclude that while the energy scale of the observed mode clearly rules out the $B_{1g}$ oxygen phonons, at least one in-plane phonon (acoustic) mode does exist at the appropriate energy.

Apart from these in-plane phonons, the STM mode might arise from inelastic co-tunneling processes[26] involving an excitation of a local vibrational mode in the intervening layers between the tip and the superconducting plane ('barrier' mode). Such `out-of-plane' phonons associated with the apical oxygen have been postulated as an alternative explanation for the BSCCO data[16]. While barrier modes in PLCCO might originate from Pr/La/Ce vibrational excitations in the layers adjacent to the $CuO_2$ planes, it is not obvious, however, how such modes would lead to our observed correlation between $\Omega(r)$ and $\Delta(r)$. Indeed, based on the idea that this correlation is significant, one recent analysis of the BSCCO STM data[27] postulates two coexisting bosonic modes, only one of which is sensitive to the superconducting gap and can be considered as a signature of the neutron resonance mode in BSCCO.

In order to bring more insight into the issue of electronic versus lattice-vibrational sources of this mode, we further explore its connection to the superconductivity energy gap. The local nature of STM spectroscopy enables a study of the relationship between the spectral properties of the local mode, and the energy scale for the onset of particle-hole excitations ($2\Delta(r)$). Figure 4a is a scatter plot showing the occurrences of the two modes $\Omega_1$ and $\Omega_2$ at a given $\Delta$ for three typical regions of different average gap sizes. It is clearly visible in the plot that the ratio $\Omega_1/2\Delta$ lies below 1 for a statistically significant fraction of the observed modes (while $\Omega_2/2\Delta$ is capped by 2, consistent with the interpretation of $\Omega_2$ as $2\Omega_1$). In figure 4b, we present the spectral lines of $d^2I/dV^2$ near the mode energy for several representative cases with different $\Omega_1/2\Delta$ ratios. Remarkably, the line-shape evolves from a symmetric sharp resonance peak-like feature at low $\Omega_1/2\Delta$ to being broad, asymmetric (like an overdamped mode) as $\Omega_1/2\Delta$ approaches and exceeds one. As shown in figure 4c, a clear anti-correlation between the sharpness of the mode and $\Omega_1/2\Delta$ is observed, providing statistical significance for the line-shape analysis in figure 4b. These findings unequivocally demonstrate the intimate connection between the bosonic mode and the quasiparticle excitations across the superconducting energy gap. We note that while our analysis of the low temperature STM data argues against the barrier modes, measurements of the normal state (T > $T_c$) DOS will provide further, more direct data for the influence of the inelastic barrier co-tunneling processes on the local tunneling spectroscopy[28].

The overall picture that finally emerges from our STM studies on PLCCO is the observation of a collective mode in the electronic excitations of the system at 10.5±2.5 meV. While we cannot rule out low-energy phonons, this mode is fully consistent with the neutron spin resonance mode, and strongly coupled to the superconducting order parameter, making it a compelling candidate boson in the scenario based on the Eliashberg[29] framework, where exchanging associated electronic (spin or charge) excitations in the normal state serves as the unconventional pairing mechanism in these materials.

**Supplementary Information** is linked to the online version of the paper at **www.nature.com/nature.**

**Acknowledgements** We thank A. V. Balatsky, E. W. Hudson, P. Richard, G. Murthy, J. Engelbrecht and J. C. Davis for discussions and comments. This work was supported by NSF and DOE.

**Author Information** Reprints and permissions information is available at www.nature.com/reprints. The authors declare no competing financial interests. Correspondence and requests for materials should be addressed to V. M. (madhavan@bc.edu).

Figure 1: Prominent low energy spectral features on PLCCO at a temperature of 5.5 K. a, A 200 Å section of a 512 Å linecut (a sequence of *dI/dV* spectra obtained along a spatial line) showing the variations in coherence peak heights and gap magnitude (Δ), defined as half the energy separation between the coherence peaks. The spectra have been offset for clarity. The gap magnitude in this linecut varies from 5 meV to 8 meV. For all spectra, V refers to sample voltage. The spectra were obtained with a junction resistance of 120 MΩ. b, A representative ±100 mV range (*dI/dV*) spectrum (200 MΩ junction resistance) illustrating the dominating V-shaped background. c, While the linecut reveals spectra that vary from ones with sharp coherence peaks to pseudogap like spectra without coherence peaks, most spectra reveal coherence peaks of varying magnitudes once the dominating V-shaped linear background is divided out. This is illustrated by the spectrum shown here from b after a linear V-shaped division. d, More examples of *dI/dV* spectra demonstrating the clearly resolved coherence peaks and modes resulting from a V-shaped division. These spectra were obtained with 200 MΩ junction resistance.

Figure 2: Gap distribution, statistics, and temperature dependence. a, A 64 X 64 pixel gap map taken on a 256 Å x 256 Å area at 5.5 K with a junction resistance of 500 MΩ. The corresponding topographic image reveals no atomic scale corrugations as is the common feature of STM images in the superconducting regions of PLCCO (see Supplementary Information) and has therefore been omitted. The patch size in these samples (defined as the area

where the gap variation is in the range ±0.5 meV) ranges from 30 Å to >100 Å. **b,** The gap distribution of the map in **a**. The average gap in this region is 6.7 ± 1.0 meV. This gap variation is much smaller than that observed in hole-doped BSCCO on a similar sized region. **c,** Multiple histograms showing gap distributions in different regions of the sample. Our STM's coarse x-y motion capabilities (±0.5 mm maximum) allow us to collect data in regions separated by larger length scales. The mean gap (for the regions represented here) ranges from 6.7 meV to 8.5 meV with a standard deviation ranging from 0.5 meV to 1 meV. Given these statistics, we conclude that gap variations in PLCCO occur on longer length scales when compared to BSCCO. The sum of these histograms is shown in blue. The average gap over all 13 maps obtained by us is 7.2 ± 1.2 meV. **d,** Temperature evolution of spatially averaged spectra. The spectra have been offset along the y-axis for clarity. A 256 Å linecut (junction resistance of 120 MΩ) was averaged at each temperature shown, from 5.5 K to 32 K. By 28 K, while the gap and coherence peaks have disappeared, the V-shaped overall background remains.

**Figure 3: Statistics of the mode observed as peaks in $d^2I/dV^2$. a,** A typical $dI/dV$ spectrum taken at 5.5 K with a junction resistance of 200 MΩ demonstrating the appearance of the modes. **b,** The same spectrum from **a** (purple) as well as it's derivative, $d^2I/dV^2$ (red). The linear V-shaped background has been divided out for clarity and the spectra are now shown only for energies greater than the Fermi energy ($E_F$). The peak in $dI/dV$ (at 7.0 meV) is the coherence peak, labeled as $\Delta_R$. The peaks in $d^2I/dV^2$ are labeled as $E_1$ and $E_2$

respectively. **c,** A histogram of the occurrences of $\Delta_R$ (purple) and the energies $E_1$ and $E_2$ (red) for a *dI/dV* map on a 64Å X 64Å area of the sample. We calculate the average gap ($\Delta_{av}$) in this region to be 7.7 ± 0.5 meV, while $E_{1av}$= 18.5 ± 1.5 meV and $E_{2av}$ ~28 meV (our cut off at 30 meV for this analysis prevents us from obtaining full statistics for $E_2$). **d,** Following convention in superconducting systems, the mode energy will be symbolized by $\Omega$ ($\Omega_i = E_i - \Delta$). $\Omega$ is calculated in two ways for each spectrum in the map: $E_1$- $\Delta_R$ (blue) with a mean of 10.7±1 meV and $E_2$-$E_1$ (green) with a mean of 10 ± 1.7 meV. These are two independent variables and the remarkable overlap between these histograms lends weight to their identification as multiples of the same mode **e,** Histogram of the mode energies $\Omega_1$ (blue) and $\Omega_2$ (pink) summed for 8 maps in different areas of the sample with gaps ranging from 6.5 meV to 8.5 meV. The mode energies were extracted from above and below $E_F$. From this data, we obtain the average mode energy $\Omega_{1av}$= 10.5 ± 2.5 meV.

**Figure 4: Variation of local mode energy and intensity with the local gap energy scale. a,** Log intensity (two dimensional histogram) plot of the occurrences of $\Omega_1$ and $\Omega_2$ plotted as a local ratio $\Omega(r)/2\Delta(r)$ against $\Delta(r)$ clearly revealing the anticorrelation between $\Omega$ and $\Delta$. Also note that $\Omega_1(r)/2\Delta(r)$ remains below 1 for a statistically significant part of the data (while $\Omega_2(r)/2\Delta(r)$ remains below 2). This demonstrates the sensitivity of the mode to the energy scale 2Δ which is also borne out by the intensity analysis in **b** and **c.** This plot includes data from three maps obtained in regions of the sample with different average gap values. **b,** Examples of $d^2I/dV^2$ spectra (from one map) for different ratios of

$\Omega_1/2\Delta$ from 0.3 to 1.03. The intensity of the mode (defined as the height of the peak in $d^2I/dV^2$ spectra) decreases and the mode gets wider in energy as $\Omega_1$ approaches $2\Delta$. Note that while both $\Omega$ and $\Delta$ can vary from spectrum to spectrum, it is the local ratio of $\Omega(r)$ to $2\Delta(r)$ that determines the intensity of the mode. This is consistent with increased damping of the mode associated with the onset of the continuum of excitations at $2\Delta(r)$. **c,** A plot of the mode intensity (reiterating the same behaviour in **b)** now for all the measured bosonic modes in a single map. Similar intensity drops with the ratio $\Omega_1/2\Delta$ were observed for maps in different regions with average gaps ranging from 6.5 meV to 8.5 meV.

Figure 1

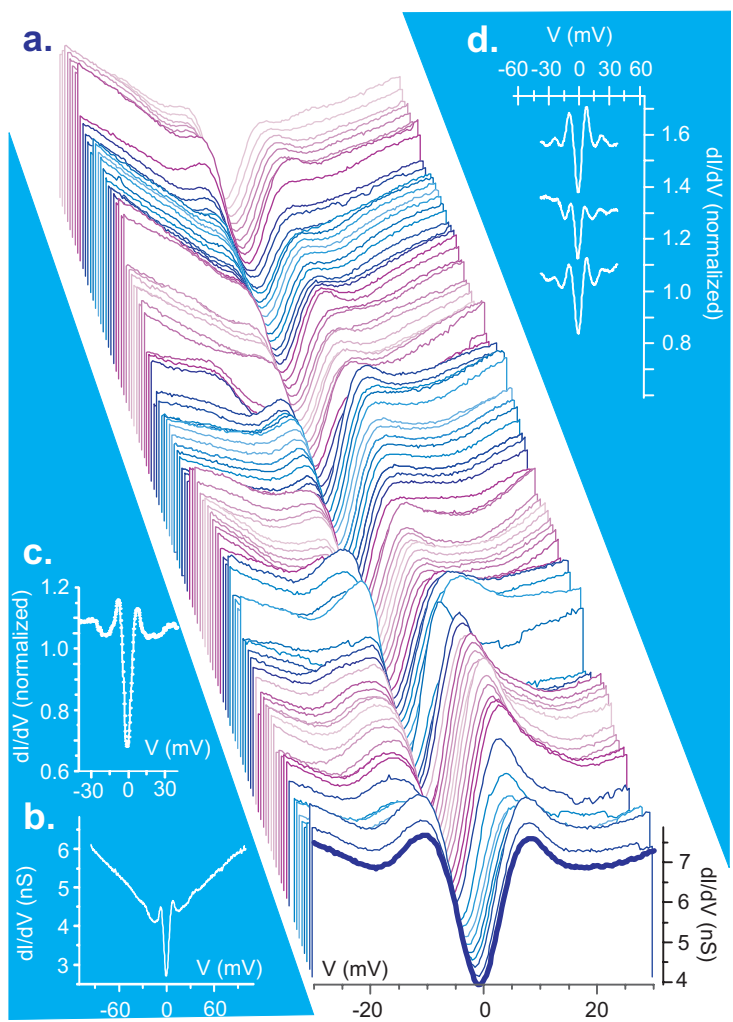

Figure 2

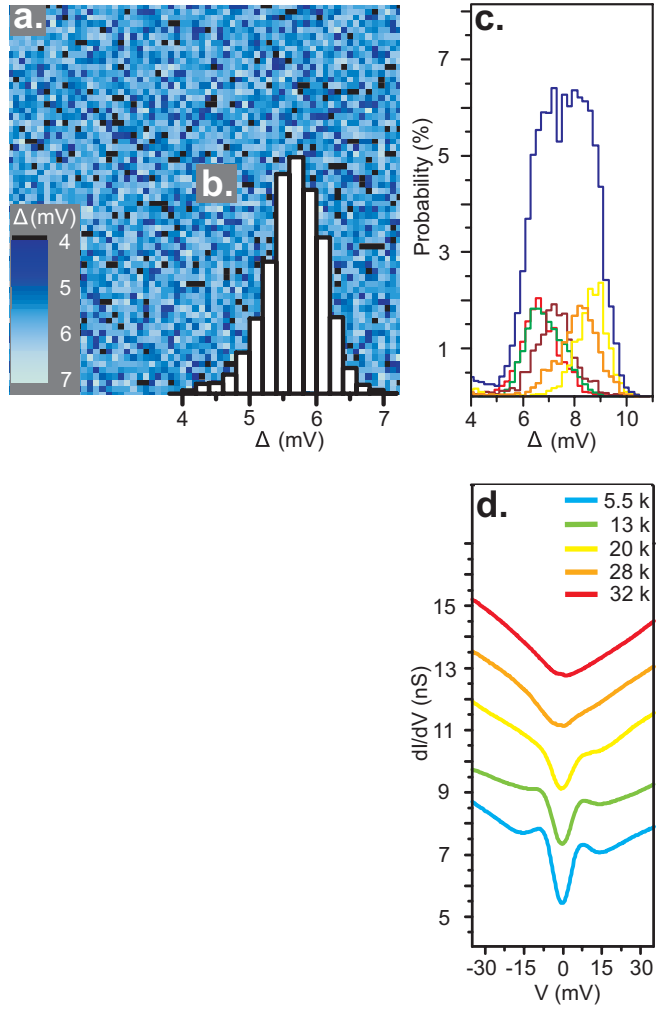

Figure 3

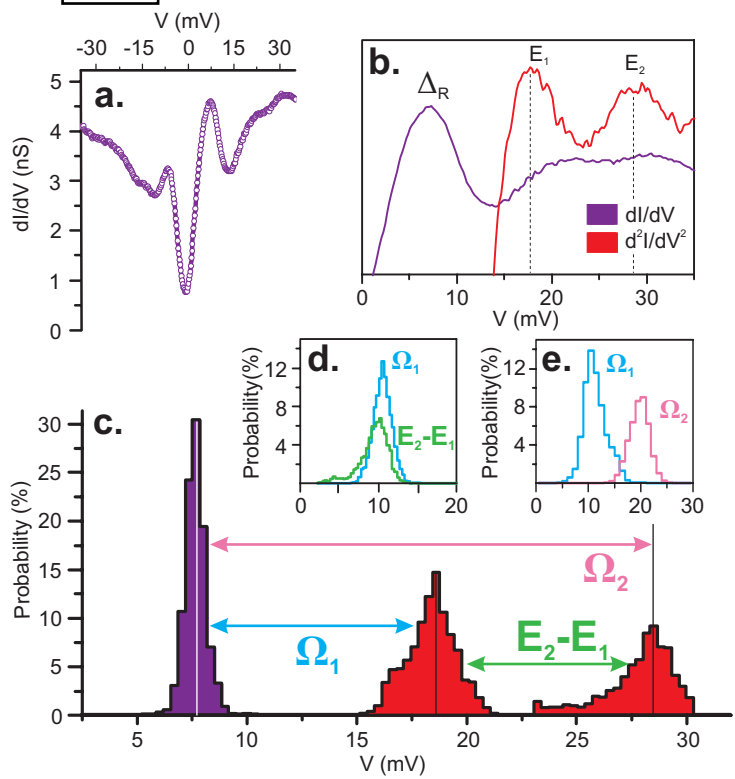

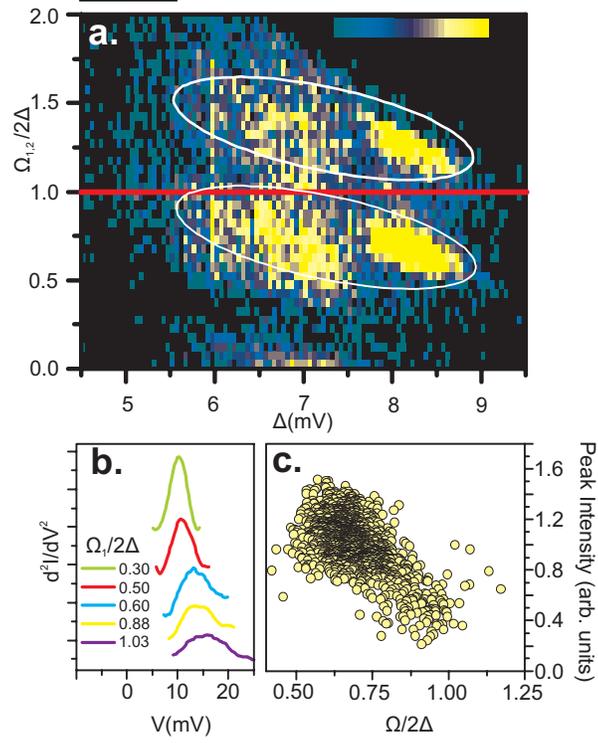

Figure 4

# *Supplement*

## Identification of a Distinct Bosonic Mode in the Electron-Doped Superconductor $Pr_{0.88}LaCe_{0.12}CuO_{4-\delta}$ by Scanning Tunneling Spectroscopy

F. C. Niestemski, S. Kunwar, S. Zhou, Shiliang Li, H. Ding, Ziqiang Wang, Pengcheng Dai, V. Madhavan

**Methods:**

The experiments were carried out with a variable temperature (2K-60K), ultra high vacuum (UHV) STM from Unisoku Co., Ltd., Japan. The machine is housed in a specially constructed sound proof room to minimize vibrations. The machine has in-situ tip and sample exchange and coarse x-y motion (± 0.5mm). This coarse motion was crucial to this experiment since it allowed us to investigate many different areas of the sample. The STM has high stability allowing us to obtain 24hr or 36 hr maps with low (few Å) drift. The low noise, high stability STM head allows us to obtain *dI/dV* spectra with good signal to noise in ~30 s (supp. fig. 3 and supp. fig. 4) allowing us to obtain large area maps in a few days.

PLCCO samples were cleaved in UHV at room temperature before direct insertion into the STM head at 5.5 K. Spectra were obtained with a modulation voltage ($V_{rms}$) between 1.0 mV - 2.4 mV.

**Summary :**

Images on the superconducting regions in PLCCO (supplementary figure 1a) do not show atomic scale features. This is not a result of blunt tips since the tips used for this study were first characterized on BSCCO and only tips that clearly revealed atoms and super-modulation on BSCCO were used to study PLCCO.

The spectra on PLCCO reveal superconducting gaps with coherence peaks as well as a dominant V-shaped background seen clearly in supp. fig. 2. Using our x-y coarse motor (±0.5mm) we were able to access different regions of the sample, well separated in space. Obtaining *dI/dV* maps on these regions allowed us to obtain statistical information on the mean and variation of the gap magnitude. Unlike hole-doped BSCCO, the gap variation on short (100Å) length scales is small (±10%). On the other hand, different regions (compare supp. fig. 2 and supp. fig. 4) of the sample reveal significantly different mean gaps values (ranging from 6.5 meV to 8.5 meV). So while the sample is inhomogeneous, the length scale of the inhomogeneity is much larger than in BSCCO.

The bosonic modes outside the gap, sometimes clearly visible (supp. fig. 3), are often masked by the V-shaped background (supp. fig. 4b). A closer look at the low energy part of the spectra shows the existence of the modes (supp. fig. 4c and supp. fig. 4d) which become much more prominent in the second derivative ($d^2I/dV^2$) of the tunnel current. The $d^2I/dV^2$ spectra show peaks and dips corresponding to step like features in the $dI/dV$ spectra (supp. fig. 5). Obtaining the $d^2I/dV^2$ spectra is therefore a crucial tool to pick out the mode energy(s).

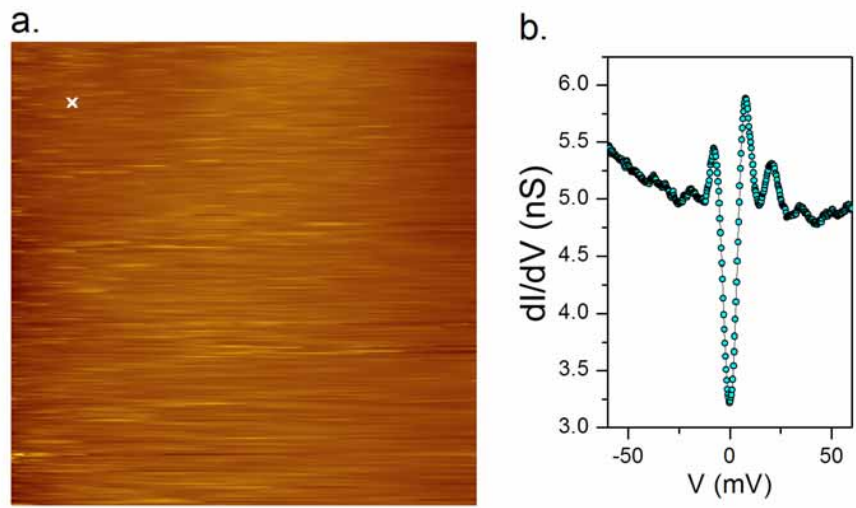

**Supplementary Figure 1: Sample topographic image and spectrum on PLCCO. a,** A 128 Å topographic image of PLCCO illustrating the lack of atomic-scale topographic features. This image was obtained with a tip that revealed atoms and super-modulation clearly on BSCCO. **b,** A representative $dI/dV$ spectrum on the spot marked `x' on the topography (V refers to sample bias in mV). The junction resistance for this spectrum is 200 MΩ.

**Supplementary Figure 2: Representative *dI/dV* spectra along a line (region 1).** Part (42 points) of a linecut (*dI/dV* spectra obtained along a line) on one region of PLCCO. The spectra were obtained with 1Å spacing. (Four spectra were replaced with a nearest neighbor average). The junction resistance for these spectra is 200 MΩ. The spectra have been offset on the y-axis for clarity in the typical waterfall plot fashion. These spectra demonstrate the ubiquitous V-shaped background observed on PLCCO. Comparing this region (average gap 8 meV) with the linecut in supp. fig. 4 (obtained with the same sample and tip, average gap 6.5 meV) reveals the importance of sampling different regions of the sample using the x-y coarse motion of our STM.

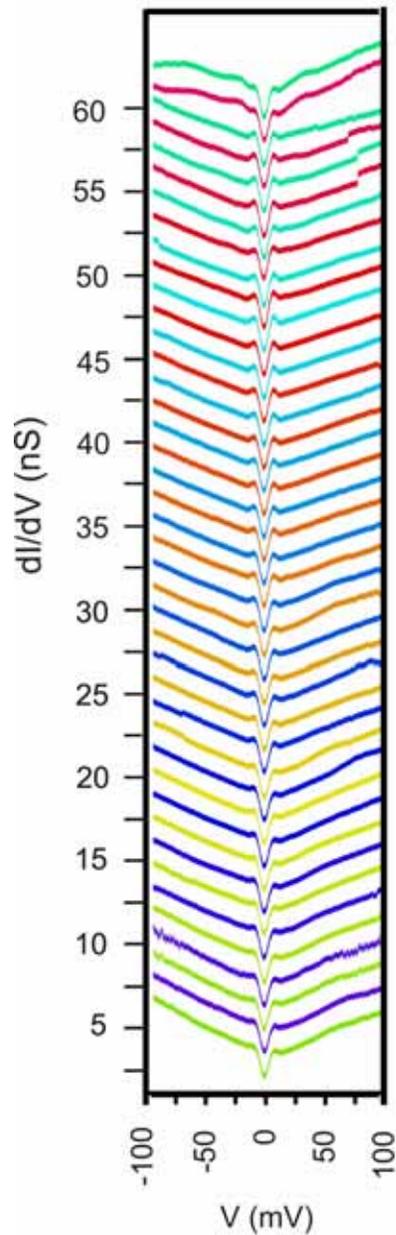

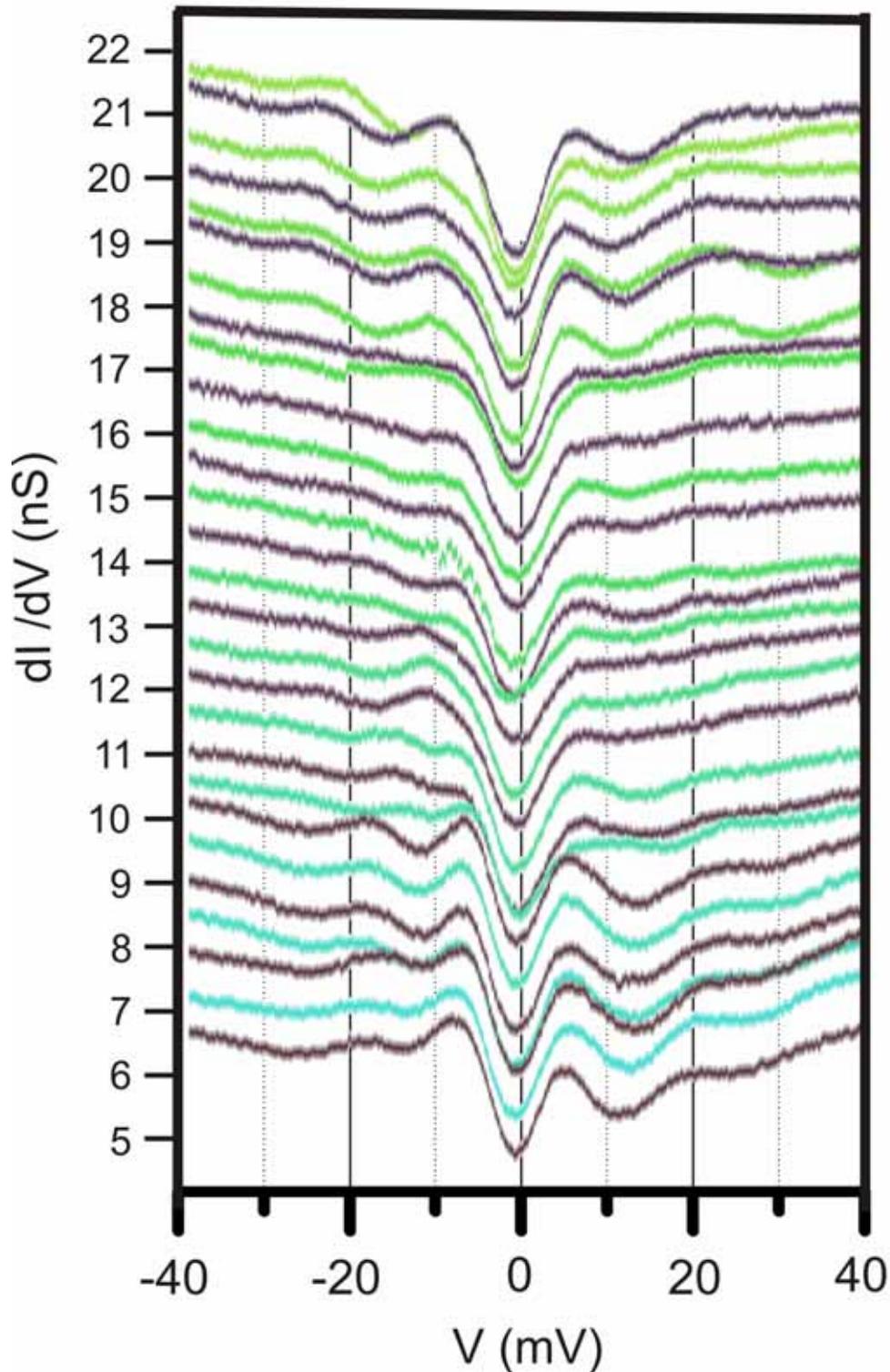

**Supplementary Figure 3: Representative *dI/dV* spectra illustrating the bosonic modes outside the gap.** Sample spectra (obtained with a tip and sample different from that in supp. fig 2 and supp. fig 4) clearly revealing the bosonic modes and the coherence peaks. It is worthwhile noting here that these spectra (and the spectra in supp. fig. 4a) were extracted from *dI/dV* maps and demonstrate the quality of the data that was used to perform the analysis in the paper. The junction resistance for these spectra is 200 MΩ.

**Supplementary Figure 4: Representative *dI/dV* spectra along a line (region 2). a,** A 82 point linecut of *dI/dV* spectra) obtained at 2 Å intervals (same tip/sample as supp. fig. 2). The spectra have been offset on the y-axis for clarity in the typical waterfall plot fashion. (10% of the spectra were replaced with a neighboring average). The junction resistance for these spectra is 200 MΩ. **b,** Zoom in of a representative spectrum marked spectrum 20. The structure outside the gap due to the bosonic mode is faintly visible. **c, and d,** The bosonic modes revealed much more clearly with the background divided out. The coherence peak position is at 5.5.meV (positive sample bias) and 6.5 meV (negative sample bias). The gap referenced mode energies ($\Omega_1 = E_1 - \Delta$) are 9.0 meV and 9.5 meV for the positive and negative bias respectively.

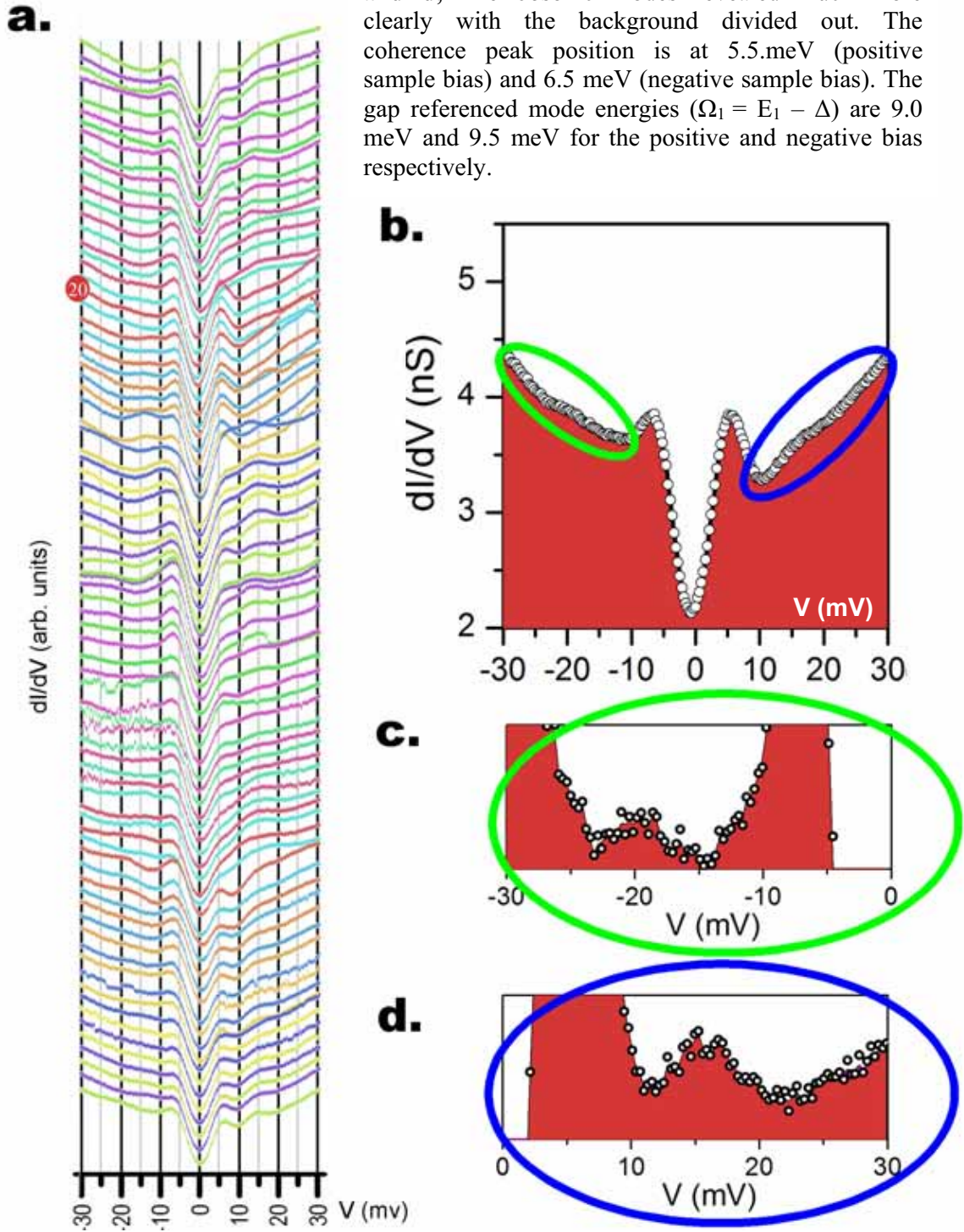

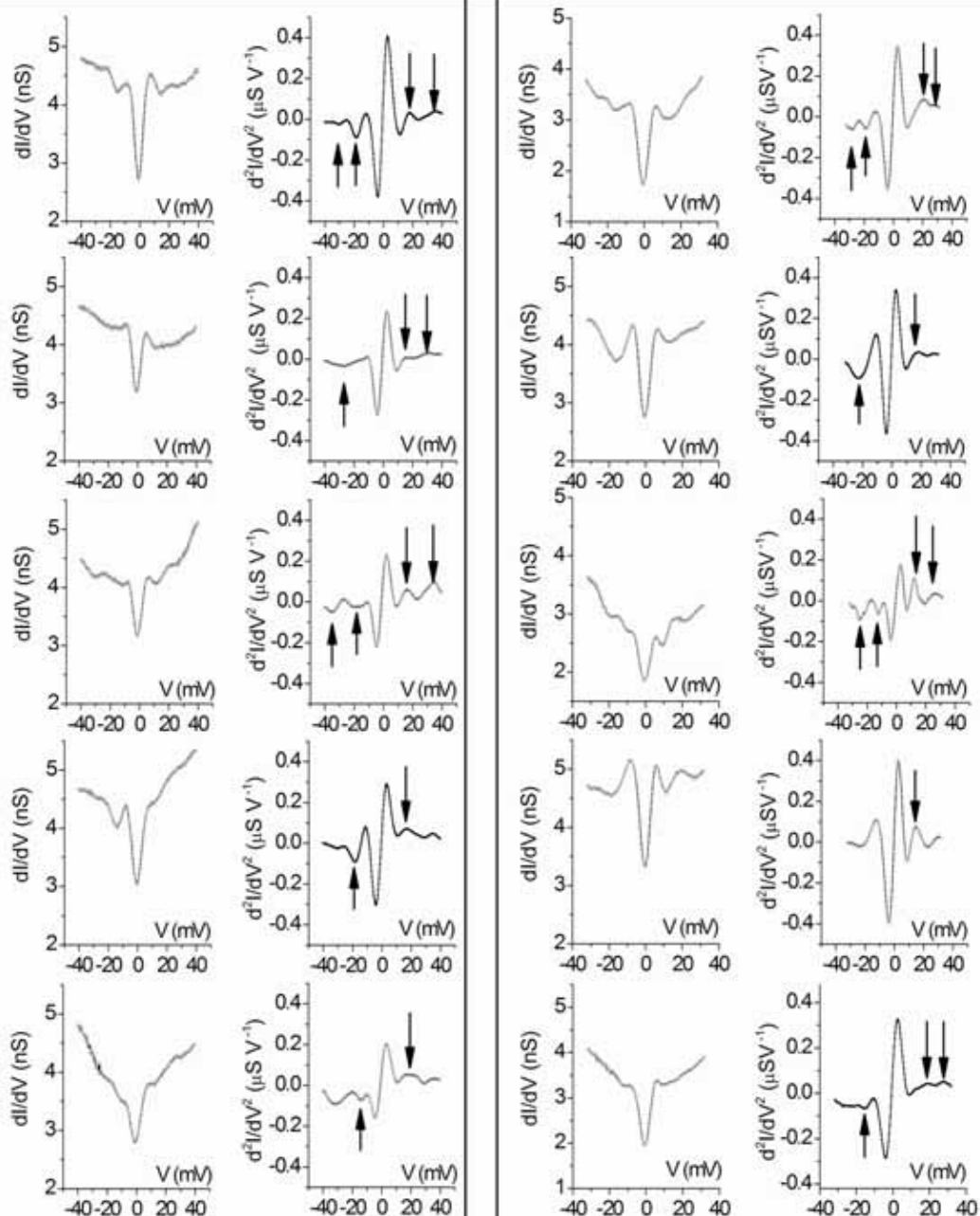

**Supplementary Figure 5: Sample *dI/dV* spectra and second derivatives *d²I/dV²*.** Spectra obtained on two different samples are displayed. The displayed spectra and derivatives were not chosen because they are `typical'. They were chosen to illustrate the different types of spectra one observes in these samples. The junction resistance for these spectra is 200 MΩ.